# Study of Capillary-based gaseous detectors

C. Iacobaeus, T. Francke, M. Danielsson, J. Ostling, V. Peskov

*Abstract*— We have studied gain vs. voltage characteristics and position resolutions of multistep capillary plates (two or three capillary plates operating in a cascade), as well as capillary plates operating in a mode when the main amplification occurs between plates or between the capillary plate and the readout plate (parallel plate amplification mode). Results of these studies demonstrated that in the parallel-plate amplification mode one can reach both high gains (>$10^5$) and good position resolutions (~100 μm) even with a single step arrangement. It offers a compact amplification structure, which can be used in many applications. For example, in preliminary tests we succeeded to combine it with a photocathode and use it as a position sensitive gaseous photomultiplier. CsI coated capillary plates could also be used as a high position resolution and high rate X-ray converter.

*Index Terms*—Capillary, GEM, PPAC, CsI cathode, Portal imaging.

## I. INTRODUCTION

Recently a great interest arose for various hole-type gas multiplication structures: capillaries [1], capillary plates (CP) [2], gaseous electron multipliers (GEM) [3]. The main fields of application for these detectors are similar to traditional gaseous detectors (wire-type or parallel-plate type), namely: detection of particles [4], X-ray and gamma imaging [5], detection of UV and visible photons [6,7]. However, hole-type detectors have several advantages over the "traditional" ones. The most important among these advantages are:
1) A possibility of charge extraction: primary electrons or electrons created due to the multiplication could be extracted from the capillary's holes and directed to another multiplication structure or to the readout plate.
2) Efficient suppression of ion and photon feedback, which could be essential in designing photosensitive detectors sensitive to visible or UV light (often called Gaseous Photo Multipliers or GPMs).

A special place among hole-type detectors belongs to CPs. This is because they are made of glass which is a well suited material for the high level of cleanliness necessary for the GPM [7].
Unfortunately, the maximum achievable gain of the hole-type detectors including CPs is usually relatively low, around $10^4$; this is marginal for the detection of single primary electrons. In order to boost the maximum achievable gain, a multistep configuration is used; the so called "cascade mode". This was especially useful for the design of GPMs: only multistep CPs and GEMs allow gains of ~$10^5$ to be reached whereas traditional detectors combined with photocathodes, sensitive to visible light, suffered from feedback even at low gains (~100) [7]. However, in other applications where the feedback is not a problem (detectors with CsI photocathodes or without any photocathode at all) the same gain could be reached with traditional gaseous detectors, for example with parallel-mesh avalanche chamber (PMAC), filled with mixtures containing low ionisation potential additives like ethylferrocene, see [8,9] and explanations in sec. IV. One of the practical problems restricting a wide range of usage of the PMAC is the necessity to use mesh-type electrodes. Difficulties associated with these are: the stretching of these meshes, bulky supporting frames and non-uniform detector response. In this work we have tried to explore the fact that CPs themselves could play the role of the mesh in some configurations. Indeed, CPs are rigid, have high quality and flatness of the surface and one can easily drift electrons through the CP. Large size (10 cm x 10 cm) CPs are commercially available. This may offer very compact detectors that will be able to operate at high gains in a parallel-plate amplification mode and at the same time, being able to suppress the feedback.
The aim of this work is to compare two modes of operation of CPs: the traditional multistep capillary amplification (CA) mode and the parallel-plate amplification (PPA) mode.
This comparison may allow us to optimize the design of the GPM as well as X-ray and γ detectors using CPs.

## II. EXPERIMENTAL SET UP

Our experimental set up consists of a gas chamber inside of which a single CP or a cascade of CPs could be installed - see Fig. 1(a, b). The chamber had two windows: one made of $CaF_2$ which is transparent for UV and the other one made of a thin beryllium foil, transparent for X-rays. Two types of CPs were used in our measurements. The first type had a sensitive area of 20 mm in diameter and a thickness of 0.8 mm. The diameter of the capillaries were 100 μm and the pitch was 126 μm. Optical transparency was 57 % [10].

The second type of CP had a thickness of 1cm, a rectangular sensitive area of 4 cm x 5 cm, and holes diameter of 0.5 mm [11]. Most of the CPs had a resistivity of > $10^{15}$ Ω between the electrodes, but we also tested several $H_2$ treated capillaries with resistivities of ~ $10^{10}$ Ω.

In the measurements aiming for the CP's gain and electron extraction efficiency we used a metallic collecting electrode placed 0.5 - 2 mm below the last CP (CP-3 in Fig. 1(a, b)).

For the position resolution measurements, the readout plate was a Pestov glass plate placed 0.5 mm or 1 mm below the last CP. The inner and outer surfaces of the Pestov glass were coated with metallic strips of 50 μm pitch. To separate the CPs from each other or from the readout plate, special ceramic or G10 rings were used. Due to the careful design of these rings we were able to apply voltages up to a few kV between the CPs. This allowed us to test the CPs in two basic modes: the CA and the PPA mode.

Since our present studies were mainly focused on the optimization of the GPM, in most of the measurements the CPs were combined with CsI photocathodes and UV light was used to create primary electrons. The CsI photocathodes were chosen for these studies because in contrast to other photocathodes (e.g. SbCs [7]), they can be exposed to air for a few minutes without a significant drop in their quantum efficiency. This feature dramatically simplified the experimental studies: the test chamber could be opened to air many times and necessary changes and modifications could be made inside.

The 20 nm thick semitransparent CsI photocathode was deposited on a $CaF_2$ disc coated with a thin layer of Cr. The disc was then placed inside the test chamber 1-2 mm above the CPs, see Fig. 1(a, b). A negative voltage was usually applied to this disc.

Depending on measurements, various UV sources were used: a pulsed $H_2$ lamp (pulse duration of a few ns) lamp, a corona discharge in air or a continuous Hg lamp.

UV light created primary electrons from the CsI photocathodes, which were then injected to the CPs. Pulses produced by the $H_2$ lamp or the corona discharge were detected by charge-sensitive amplifiers. The photocurrent produced by the Hg lamp was measured with a Keithley picoampermeter. A system of calibrated neutral density filters was often used to attenuate the UV beam on several orders of magnitude.

Experiments with the CPs were performed in various Ar, Xe and Kr-based gas mixtures with 5-20% of various quenchers: $CH_4$, $CO_2$ or isobutane at a total pressure of 1atm.

In some measurements X-rays with energies between 6 and 20 keV were used. They were produced either by a $^{55}$Fe source or by a Kevex or a Philips tube. In these measurements the $CaF_2$ disc was placed at a distance of 1 cm above the CP-1 top and this space was used as an absorption/drift region (see Fig. 1(a, b)).

Alpha particles ($^{241}$Am) were used in some studies. The alpha source was installed 3 cm above the CP-1 on a special mesh which at the same time served as a drift electrode. The alfa source faced into the drift region without any collimation except its flat mounting media and the metallic mesh drift electrode.

### III. MEASUREMENT PROCEDURES AND RESULTS
*III.1 MEASUREMENTS WITH UV SOURCES*
*III.1-1.MEASUREMENTS IN CA MODE*

The gain of the CPs and the charge extraction efficiency were measured by the following procedure. We connected all electrodes of the CP-1 and measured the charge signal. The signal was produced by the electrons extracted from the CsI photocathode as a function of the applied voltage between the $CaF_2$ disc and the CP-1. Typically, this signal increases sharply with the voltage and then reaches a saturation value. The saturation value is determined by the convolution of the UV source emitting spectra with the CsI quantum efficiency. The absolute value of the signal (we call it the "injection signal") $Q_{inject}$ could be changed in a controllable way, on the order of magnitudes, with the help of neutral density filters. After adjusting the injection signal to a desired value, we performed a series of measurements with it. This allowed us to determine the gain of the capillaries and the charge extraction efficiency from them. For example, to measure the gain of the CP-1 as a function of applied voltage, we connected the top electrode of the CP-1 to the HV power supply, and kept the other CPs electrodes (CP-1 bottom and CP-2) connected together. Then at a fixed voltage drop between the $CaF_2$ disc and the cathode of the CP-1 we measured the signal $Q_{meas.}$ on the CP-1 bottom as a function of the voltage drop over the CP-1.

The measured signal was in fact the sum of two signals: one induced signal through the CP parasitic capacitance (capacitance between CP's electrodes) Q induced and a real signal due to the electron extraction and/or multiplication $Q_{mult.}$. Hence

$Q_{meas.} = Q_{induced} + Q_{mult.}$ (see [12] for more details).

For our CPs the signal $Q_{induced}$ was const=0.1 $Q_{inject}$, so in gain measurements it could be neglected. If we define K= $Q_{meas.}/Q_{inject}$ then the gain A is A=K at K>1. In a similar procedure we measured overall gains of double and triple CPs. Some of these measurements are presented in Fig. 2. One can see that gains up to $10^4$ were possible to achieve with a single CP and up to $3*10^5$ with triple CPs.

Usually, for an equal voltage drop over the CP-1 and the CP-2 (let's call them "CP-1 voltage" and "CP-2 voltage"), gains up to 10 times higher were safely (without any sparks) achieved with two CPs operating in tandem, compared to a single one, see Fig. 2. However, stable operations at gains of more than $4 \cdot 10^4$ were difficult to achieve due to sparks and their propagation from one CP to another. Brief studies showed that the feature of this phenomenon was very similar to the one which occurs in muti step GEMs, see [13].

Overall gains of $\sim 10^5$ were achieved only when the CP-2 voltage was ~50 V higher than CP-1 voltage and the CP-3 voltage was ~50 V higher than on the CP-2 voltage. Some results of the gain measurements at such voltage settings are indicated in Fig. 2. At this condition it was possible to observe pulses produced by single photoelectrons extracted from the CsI photocathode.

In a similar procedure, to the one described above, we were also able to measure the electron extraction efficiency from one CP to another. For example, to measure the electron extraction efficiency from CP-1 to the CP-2, all electrodes of CP-2 were connected together to a charge sensitive amplifier (see Fig. 1(a)). We measured the signal from this amplifier as a function of the voltage drop between the CP-1 bottom and the CP-2 (at a constant CP-1 voltage). As an example, Fig. 3 shows some results both for thin and thick CPs (marked as CP-1 in Fig 1(a)) obtained in Ar + 10%CH$_4$ at 1 atm. One can see that in both cases it was possible to fully extract the primary electrons drifted through the capillaries. The only difference was in the CP-voltage value at which a 100% transfer efficiency was reached.

For the position resolution measurements we replaced the metallic collecting electrode with the Pestov glass plate covered with metallic strips. This readout plate was placed 0.5 mm below the CP-3. A screen with a slit of 30 μm was attached to an alignment table that was put in close contact with the CaF$_2$ disc. The alignment table was controlled by stepper motors and allowed not only to rotate the slit, but also to move it in one direction with micron accuracy. With the help of this table the slit was aligned to be exactly parallel to the readout strips. The corona discharge source was moved far away enough (~20 cm) from the screen to avoid any geometric smearing effects. The small size of the corona discharge (<1mm) allowed us to perform very accurate position resolution measurements. The corona discharge could operate both in continuous and in a pulsed mode. Fifteen central strips facing the slit were connected to individual charge sensitive amplifiers. This allowed us to measure the induced charge zone on the readout strips. In some measurements the slit was moved perpendicular to the readout strips and a signal from a single readout strip was measured as a function of the slit position. Some results are presented in Fig. 4. In these measurements the slit was oriented in parallel with one of the hole rows and aligned to the hole centers. As one can see, the FWHM of this distribution is about 0.5 mm. With a single CA (see Fig. 1b) it was ~ 350 μm.

Note that the size of the induced charge is directly related to the position resolution. The size of the induced charge signal becomes crucial at high rate operation, because in this case: 1) a pileup of pulses is possible, 2) it is difficult to apply a centre of gravity method online, or any other analog interpolation method.

### III.1-2 MEASUREMENTS IN PPA MODE

#### A. PULSED UV SOURCE

As was mentioned in the introduction, the main goal of this work was to compare the conventional CA mode with the PPA mode. For measurements of gain and position resolutions in the PPA mode, the set up shown in Fig. 1(b) was used. The CaF$_2$ window and the readout plate were placed at distances of 2 and 0.5 mm respectively from the CP.

Gain measurements were performed with the corona discharge operating either in a current or in a pulsed mode. Photoelectrons extracted from the semitransparent CsI photocathode were directed by the electric field towards the CP. The CP voltage was kept at about ~1 kV to ensure full transparency for the primary electrons drifted through the capillaries (without any multiplication inside the capillaries). The multiplication occurred only in between the anode of the CP and the readout plate, where gap voltages up to 3 kV were applied. The procedures of the gain and the position resolution measurements were similar to the ones described in section III.1-1. Some of our results are presented in Fig. 5. As one can see, gains over $10^5$ were easily achieved in a single-step multiplication mode. In some gases, for example Ar + isobutane, even a single electron peak was possible to observe, which is essential for the efficient detection of UV and visible photons. No feedback pulses were observed up to gains of $10^5$. The FWHM of the induced charge zone on the readout strips was extremely narrow, ~100 μm. Note that the exact value of the FWHM critically depended on the alignment. It had the least value when the slit image was projected on to one of the readout strips and was parallel to this strip and to one of the CP's hole row and aligned exactly at the hole centers. However, when the slit was aligned in between two strips or was not parallel to the holes row, the FWHM increased up to a factor 2.5 - 3. When the length of the slit was reduced to 100 μm the FWHM was varied, depending on the alignment, between 150 and 300 μm. One may consider this charge distribution as a kind of "point spread function" for this particular alignment. Thus due to the excellent alignment set up we have reached the limit where the FWHM was sensitive to the alignment and at the best alignment was determined by the hole diameter and the trip pitch.

We also performed some measurements in a combined CA and PPA mode. For this test the CP operated at a voltage drop of 1.3 kV to ensure an avalanche gain inside the capillaries of ~5. Some results are presented in Fig. 4 and 5. One can see that overall gains close to $10^6$ and position resolutions of ~100 μm were safely achieved at a proper alignment.

The general conclusion one can derive from these measurements is that the PPA mode may offer a simple and compact detector design able to operate at high gains and with a potentially very narrow induced charge distribution at the readout plate.

## B. MEASUREMENTS WITH CONTINUOUS UV SOURCES

It is known that PMAC can operate at extremely high counting rates, see for example [14]. Thus it will be important to investigate what limit, on the rate characteristics of the detector operating in PPA mode, will be imposed by the CP used as a cathode, see Fig. 1b. For this reason we performed a series of measurements in a current mode.

First we connected the CP's electrodes together with the readout plate, and measured the current from them as a function of the applied voltage between the $CaF_2$ disc and the CP. This current increased with the voltage V and then at V > 1 kV reached the saturation value $I_{inject}$. This saturation value was determined by the convolution of the Hg lamp emitting spectra, with the CsI quantum efficiency.

Then we measured the CP's gain and extraction efficiency as a function of the voltage applied to its electrodes at various light intensities (or, more precisely, at various currents from the CP). For this test we connected the CP bottom electrode with the collecting plate, and measured the current $I_{cp}$ as a function of the voltage drop over the CP electrodes. The voltage drop between the $CaF_2$ disc and the CP was fixed at, $\Delta V= 1.5kV$, see Fig. 3. The ratio of $I_{cp} / I_{inject}$ gave us the gain of the CP operating in a current mode, or in other words at high counting rate. Some of the results of these measurements are presented in Fig. 6. To avoid any possible risk of breakdowns, which may fatally damage our picoamperemeter, such measurements were performed only at low gains of the CPs (below 50). At this condition one can see that only at a current of > 100 pA a conventional CP starts to exhibit some signs of charging up effects. In a similar procedure we measured the "extracted" current from the CP ($H_2$ treated in this case) to the readout electrodes at a voltage drop between the CP bottom and the readout electrode of 2 kV. Results are presented in Fig 6. One can conclude that in the case of the $H_2$ treated CPs, no charging up effect was observed up to a current of 500 pA. At $I_{inject}$ =3 pA the maximum achievable gain of the detector in the PPA mode was $\sim 2 \cdot 10^3$, which is remarkable for this counting rate (see [14] for more details).

## III.2 COMPLIMENTARY MEASUREMENTS WITH X-RAYS AND ALPHAS

In several of the early works [2, 15, 16] CPs were used for detection of X-rays photons. In these works X-rays were absorbed in a gas drift volume and the created primary electrons were amplified inside the capillaries. In recent works (see [5, 17] and references within) CPs coated with CsI layers were used also as X-ray converters. Primary electrons extracted from the CsI layer drifted through capillaries without any amplification, were extracted from the capillary holes and directed to an independent amplification structure. As was shown in [18], the maximum achievable gain of CPs and PPACs for UV and X-ray photons, may be very different. Thus it is important to compare the maximum achievable gain reached with UV and with X-rays for both the CA and the PPA modes of operation.

For measurements in the CA mode we used a set up as shown in Fig. 1(a, b), but with the $CaF_2$ disk replaced by a mesh. The space between the drift mesh and the CP-1 (about 1 cm) was used as an absorption region for X-rays. For measurements in PPA mode the set up was similar to the one shown in Fig. 1b, but without any $CaF_2$ disc (it was replaced with a mesh). The drift space was reduced to 2 mm. In some comparative studies CP coated with CsI were used as well (see [5] for more details). In this case the drift mesh was put in direct contact with CP-1. In all the measurements the Pestov-glass plate with metallic strips was used.

For gain measurements, all strips were connected to one single charge sensitive amplifier. Fig. 7 shows a typical pulse shape of the charge signal measured with 20 keV X-rays in a CA mode.

For position resolution measurements, an X-ray beam collimated to 30 μm was used. To achieve the best result the alignment of the X-ray beam was the same as described in sec. III.1-1. As in the previous case, 15 strips facing the collimated beam were connected separately to 15 charge-sensitive amplifiers. In some control measurements the collimated X-ray beam was moved in the direction perpendicular to the strips and the induced charge signal, was measured from a single strip.

The results obtained with X-rays were qualitatively similar to the ones obtained with the UV (see Fig. 4). Fig. 2 and 5 shows gains vs. voltage curves for two modes: the CA and the PPA. As one can see, gains up to $\sim 10^5$ were possible to achieve in single step PPA mode or cascaded CA. Note that the gains reached with X-rays were almost the same as in the case of UV photons.

Fig. 4 shows some results of the position resolution measurements. One can see that in the case of the PPA mode the size of the induced charge zone was much narrower that in the case of the CA mode.

Final measurements were done with an alpha source to check the reaction of CPs on heavily ionizing charged particles. Results are presented in Fig. 2. Much lower gains were achieved with alpha particles, which is expected due to the limit imposed by the Raether criteria (see [14] and next section).

## IV. DISCUSSION

Table 1 summarizes the most important results presented in this paper. These results clearly demonstrate that the PPA mode offers a higher gain in one step and potentially a better position resolution.

As was shown in [18], the maximum achievable gain of usual micropattern detectors (not combined with photocathodes) is determined by the Raether limit:

$$A_r n_0 = \text{const} \ (10^6\text{-}10^8 \text{ electrons}),$$

where $A_r$ is the gas gain, $n_0$ the number of primary electrons triggering the avalanche. The exact value of the Raether limit depends on the detector's geometry and $n_0$ [18]. As was established earlier [18], the Raether limit for PMAC is higher than for CPs. On the other hand one could expect, that if the cathode of the PPAC is made of a CP, it will have the same gain limit as a usual PPAC. Note that this was confirmed experimentally in this work, see Fig. 5.

It is known also [7] that if micropattern gaseous detectors are combined with photocathodes the maximum gain may be limited by photon and ion feedback processes. In this case the maximum achievable gain $A_f$ is determined by the equation:

$$A_f \cdot \gamma = 1,$$

where $\gamma$ is the probability of secondary electrons appearing due to the feedback mechanisms. It is important to note that $\gamma$ depends on the value of the electric field near the cathode and on the ionization potential of the gas [19]. For example it has the lowest value for gas mixtures with small ionization potential, and this is why one could reach very high gains without feedback problems in mixtures with etylferrocene (see sec.I).

Usually $A_f < A_r$ - the maximum gain achieved without photocathodes. However, in the case of PPA mode, due to efficient feedback suppression by the capillaries, $A_f \sim A_r$ even when the detector was combined with the CsI photocathode.

This approach may offer a simple and compact design of detectors for UV and X-ray photons. Based on the obtained results, one can also expect that in the case of CPs coupled with a photocathode sensitive to visible light a combination of CA and PPA modes looks like a very attractive way to reach high gains [20].

Besides the maximum achievable gain, the other important issue is the position resolution of the CP-based gaseous detectors. In the case of a bunch of single primary electrons (UV light) and a proper alignment, the size of the induced charge zone was significantly smaller for the PPA mode. This was because there was no multiplication inside the capillaries and avalanches developed in the parallel plate structure only. Since the charge in the avalanche grew exponentially with the distance from the anode, the maximum induced signal was when the avalanche was close to the readout strips. As a result, the size of the induced charge zone was significantly smaller than in the case of an electron cloud drifting to the readout plate without any multiplication (CA mode).

In the case of X-rays, the size of the induced charge zone is mostly determined by the range of photoelectrons. The last one, depending on photon energy, may be rather large and reach several mm. In this case, to improve the position resolution, analogue interpolation methods should be used. Unfortunately there could be some difficulties, to implement this method at high counting rates (close to or above $10^5$ Hz/mm$^2$) due to signal pile-ups and on line electronics speed limitations. On the other hand, in some applications where the energy resolution is not very important, one can use CP as a converter of X-ray photons to primary electrons. Two options were experimentally tested: conversion of X-rays in the gas volume inside the capillaries and conversion in capillary surfaces coated with a CsI layer. In both cases a position resolution of 100 μm was easy to achieve.

## V. CONCLUSIONS AND OUTLOOK

Performed studies allowed us to optimize the design of GPM with CsI photocathodes as well as X-ray detectors using CPs as X-ray converters. In particular, it was found that PPA mode offers both high gas gains and high position resolutions. Special benefits may exist when CPs operate in a combined mode of both the CA and the PPA.

Results of these studies could also be useful for the optimization of other CP–based detectors, for example: a portal imaging device [5, 21], X-ray and γ polarimeters.

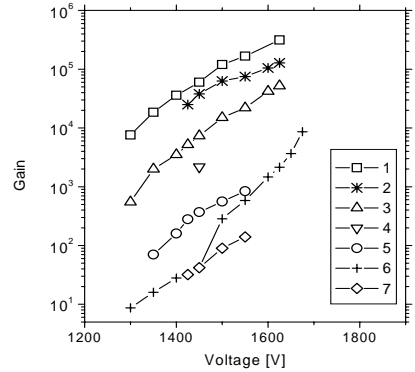

Fig. 2. Gain vs. voltage for CPs operating in CA mode. Results obtained with the UV and the CPs operating in a CA mode:
1. Overall gain of 3 CPs at the CP-3 voltage being 50 V higher than on the CP-2 and the CP-2 voltage being 50V higher than on the CP-1.
3. Overall gain of 2 CPs at the CP-2 voltage being 50 V higher than on the CP-1, 4. Gain of 2 CPs at equal voltages over the CP-1 and the CP-2; 6. Gain of a single CP;
Results with X-rays: 2. Overall gain of 3 CPs measured at the same voltage setting as the "1" curve.
Results with alphas: 5. Overall gains measured with double CPs. 7. Gains measured with a single CP.

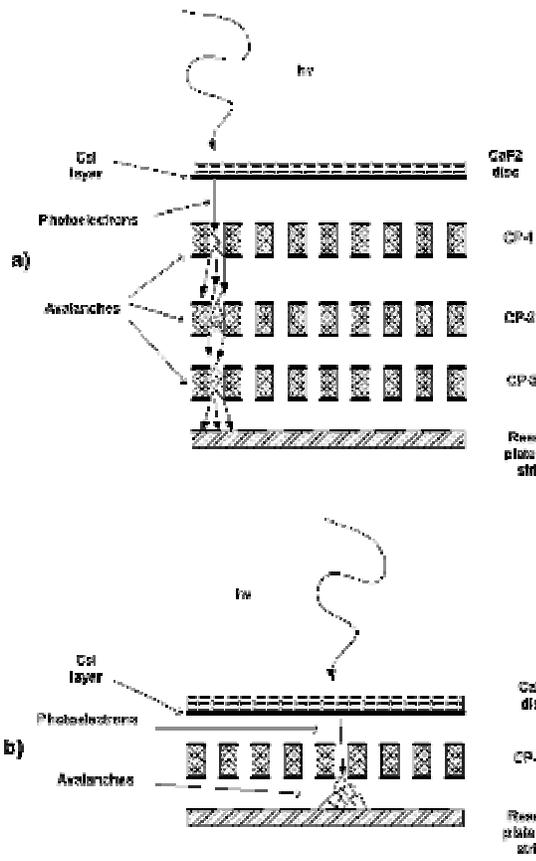

Fig. 1 (a) Set up for studies of CP's operating in CA mode. (b) Set up for studies of CPs operating in a PPA mode.

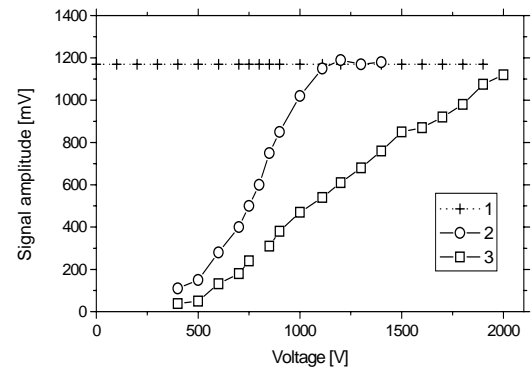

Fig. 3. Results of measurements of the charge extraction from the CP-1 to the CP-2 as a function of the voltage in between the CPs. Signals on CP-1 bottom. 2. Signal from the CP-2 top (0.8 mm thick). 3. Signal measured at the same conditions as for the 1 cm thick CP.

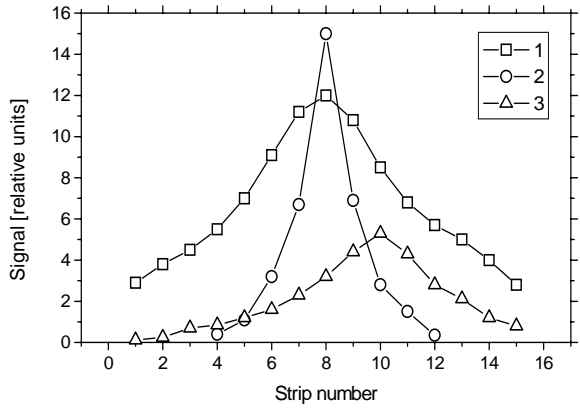

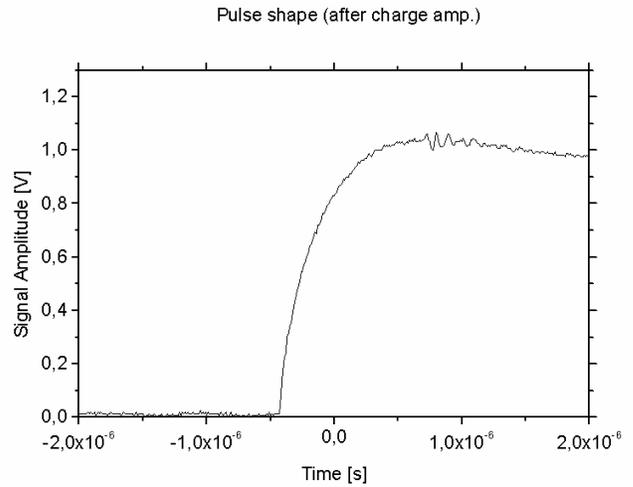

Fig. 4. The best results obtained with induced charge profile measured with UV and X rays. Collimated UV:
1. Three CPs operating in a CA mode. 2. A single CP operating in the PPA mode (note that almost the same profile was measured in the case of the combined CA and PPA modes). 3. Results obtained with 6 keV x-rays and the CP operating in the PPA mode.

treated CP at $I_{inject}$=10 pA and at voltage drop between the CP and the readout plate of 2kV.

Fig. 7. A typical pulse shape measures with CP operating in the CA mode.

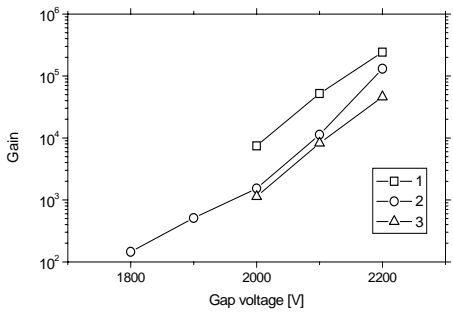

Fig. 5. Gain vs. gap voltage for CPs operating in PPA mode and in a combined mode of CA and PPA, measured with UV and X-rays. 1. Gains achieved in combined CA and PPA modes, UV. 2. Gain of a single CP operating in a PPA mode, UV. 3. Gain of single CP operating in the PPA mode, X-rays.

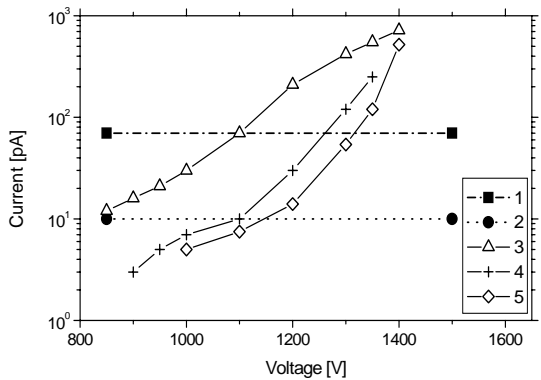

Fig. 6. Results of measurements in a current mode vs. voltage drop over the CP for two values of the $I_{inject}$: 70 pA (plot 1) and 10 pA (plot 2). 3. Current measured on the CP bottom for at $I_{inject}$=70 pA. 4. Current on the CP bottom at $I_{inject}$=10 pA. 5. Current measured on the readout plate in the case of a $H_2$

Table I

| Set ups: | Max. achiev. gain with pulsed UV | Max. achiev. gain with X-rays | Max. achiev. gain with alphas | FWHM (μm) with pulsed UV | FWHM (μm) with X-rays |
|---|---|---|---|---|---|
| Fig. 1a double CPs | $4*10^4$ | | $10^3$ | | |
| Fig. 1a triple CPs | $3*10^5$ | $10^5$ | | ~500 | |
| Fig. 1b CA mode | $10^4$ | | $1.5*10^2$ | 350 | |
| Fig. 1b PPA mode | $10^5$ | $5*10^4$ | | 100-300 (dependins on alignment) | ~300 |
| Fig. 1b CA+PPA mode | $3*10^5$ | | | 100-300 (depends on alignment) | ~300 |